\documentclass[aps,prb,twocolumn,groupedaddress,showpacs]{revtex4}

\usepackage{graphicx}
\usepackage{dcolumn}
\usepackage{bm}
\begin{document}


\title{Biquadratic antisymmetric exchange and the magnetic phase diagram of magnetoelectric CuFeO$_2$}


\author{M.L. Plumer}
\affiliation{Department of Physics and Physical Oceanography, Memorial University, St. John's, Newfoundland, Canada, A1B 3X7}

\date{\today}

\begin{abstract}
Biquadratic {\it antisymmetric} exchange terms of the form  $  -  [C_{ij} e^{\alpha}_{ij}({\bf s}_i\times{\bf s}_j)_z]^2$, where ${\bf e}_{ij}$ is the unit vector connecting sites $i$ and $j$ and $\alpha = x,y$, due partially to magnetoelectric coupling effects, are shown to be responsible for the spin-flop helical phase in CuFeO$_2$ at low magnetic field and temperature. Usual biquadratic {\it symmetric} exchange, likely due to magnetoelastic coupling, is found to support the stability of axial magnetic states at higher fields in this nearly-Heisenberg like stacked triangular antiferromagnet.   A model Hamiltonian which also includes substantial interplane and higher-neighbor intraplane exchange interactions, reproduces the unique series of observed commensurate and incommensurate periodicity phases with increasing applied magnetic field in this highly frustrated system. The magnetic field-temperature phase diagram is discussed in terms of a Landau-type free energy. 
\end{abstract}


\maketitle
\section{Introduction}
The unusual magnetic properties of the stacked triangular antiferromagnet (STAF) CuFeO$_2$ have been the subject of considerable investigation and speculation over the past several decades, especially since the recent discovery of an unconventional magnetoelectric (ME) coupling induced by relatively weak magnetic field. \cite{kimura06,seki06}  The focus of the present work is on the surprising sequence of magnetic states which occur at low temperature with increasing field strength applied along the hexagonal c-axis. \cite{petrenko00,mitsuda00,kimura06,terada06,nakajima07}  These can be characterized with the following spin polarization vectors ${\bf S}$ and in-plane periodicities: linearly polarized ${\bf S} || {\bf c}$  period-4 (P4), helically polarized ${\bf S} \perp {\bf c}$ incommensurate (IC), linearly polarized ${\bf S} || {\bf c}$  period-5 (P5), axial and canted ${\bf S}$ period-3 (P3) phases. Although the rhombohedral R$\bar{3}$m crystal symmetry present at temperatures above the onset of linearly polarized ${\bf S} || {\bf c}$ IC magnetic order in zero field at $T_{N1}$=14~K allows for axial magnetic anisotropy, usual spin-orbit coupling is absent in the S=5/2, L=0 magnetic state of the $Fe^{3+}$ ions. \cite{petrenko05,whangbo06} (The extent to which Hund's rules apply to this semiconductor with non-trivial electronic structure \cite{galakhov97} may be questioned.) Anisotropy may thus be expected to be small so that the origin of stability of the axial phases is not obvious. The existence of such a wide variety of magnetic states with different periodicities in this compound is in contrast with the other weakly axial STAF's such as CsNiCl$_3$ which show only period-3 phases and a typical spin-flop transition. \cite{plumer88}  Neutron scattering experiments also indicate a period-two modulation of the magnetic order in CuFeO$_2$ along the c-axis of the corresponding hexagonal unit cell suggesting substantial antiferromagnetic (AF) interplane coupling. \cite{petrenko00,nakajima07}   Such a complex phase diagram defies explanation through a model Hamiltonian  based on conventional magnetic interactions. 

Remarkably, some important aspects of the magnetic ordering (e.g., stability of P4 and P5 states) have been accounted for by a model 2D triangular Ising AF with very large second and third-neighbor exchange interactions, $J_2/J_1$=0.45 and $J_3/J_1$=0.75, respectively, where $J_1$ is the nearest-neighbor (NN) exchange. \cite{mekata93,ajiro94,fukuda98}  The physical origin of these unusually long-ranged exchange effects is unknown, \cite{petrenko05,whangbo06} but there is evidence from perturbed angular correlation measurements that $J_2$ and $J_3$ are substantial. \cite{uhrmacher96}  Analysis of the model proposed here demonstrates the importance of inter-plane coupling, $J'$, \cite{petrenko05,terada07} for stabilizing the P4 state with smaller values of $J_2$ and $J_3$, an effect which has been found in other frustrated systems. \cite{plumer92} 

It has also been emphasized that spin-lattice coupling is an important effect in CuFeO$_2$, as observed in magnetostriction \cite{kimura06} and in the pressure dependence of $T_N$. \cite{xu04} Such interactions are also believed to be relevant for the structural phase transition to monoclinic $C2/m$ symmetry which occurs as the temperature is lowered to about 11~K. It is near this temperature $T_{N2}$ that a discontinuous magnetic phase transition from the linear IC phase to the ground state P4 ordered structure occurs. \cite{ye06}  It is of interest to note that although the type of magnetic order is quite different, a similar $C/2m$ - $R \bar{3}m$ structural phase transition in solid $O_2$ has been shown to be driven by magnetoelastic coupling. \cite{freiman04}  
Magnetoelastic coupling is known to give rise to biquadratic {\it symmetric} exchange terms of the form  
$\mathcal{H}_G = -\sum_{ij} G({\bf r}_{ij})[{\bf s}_i\cdot {\bf s}_j]^2$, where ${\bf r}_{ij} = {\bf r}_j-{\bf r}_i$, \cite{kittel} and has been related to the P4-IC phase transition in Ho. \cite{plumer91a}  This type of coupling can also arise from higher-order effects within the Hubbard model.\cite{takahashi77} Such coupling terms favor linearly polarized magnetic order \cite{walker} and are shown here to be partly responsible for the stability of the modulated axial phases with ${\bf s} || {\bf H}$ in CuFeO$_2$ at high fields even in the absence strong anisotropy.   
 
\section{Antisymmetric Exchange} 
 
Key to understanding the magnetic phase diagram of CuFeO$_2$ is the occurrence of an induced electric polarization ${\bf P}$ in the IC spin-flop phase. \cite{kimura06}  This type of ME effect \cite{fiebig05} is atypical due to the fact that the delafossite crystal space group R$\bar{3}$m contains inversion symmetry.  Coupling between noncollinear spins at sites i and j and the electric polarization can be shown to be compatible with R$\bar{3}$m symmetry if it also involves the lattice vector ${\bf r}_{ij}$ and is of the form \cite{shiratori80,katsura07}
\begin{eqnarray}
\mathcal{H}_C = \frac{1}{2} \sum_{ij} C({\bf r}_{ij})({\bf P}_{ij} \times {\bf e}_{ij}) \cdot ({\bf s}_i\times{\bf s}_j)_z
\end{eqnarray}
where ${\bf e}_{ij} = {\bf r}_{ij}/r_{ij}$, $C({-\bf r}) = C({\bf r)}$ and the subscript $z$ indicates the ${\bf \hat{z}}$ component, parallel to the hexagonal $c$ axis. Such coupling involves only components ${\bf P} \perp {\bf \hat z}$ and should be present in all crystal systems which contain a symmetry plane where the spin Hamiltonian is isotropic in {\bf s} at second order.   It has also been used as a mechanism for electric-field induced chirality selection. \cite{plumer91b}

Following Katsura {\it et al.} \cite{katsura07} and Bergman {\it et al.} \cite{bergman}, consider now adding to the Hamiltonian the lowest order contribution in ${\bf P}$ as a NN sum $\mathcal{H}_P = \frac{1}{2} A_P \sum_{<ij>} P_{ij}^2$.  Elimination of ${\bf P}$ by minimizing $\mathcal{H}_{CP} = \mathcal{H}_C + \mathcal{H}_P$ results in a new type of {\it biquadratic antisymmetric exchange} interaction of the form 
\begin{eqnarray}
\mathcal{H}_{CP} = -\frac{1}{8A_P} \sum_\alpha  \sum_{<ij>} \left[ C({\bf r}_{ij}) e^{\alpha}_{ij} ({\bf s}_i\times{\bf s}_j) \cdot {\bf \hat z}\right]^2 
\end{eqnarray}
where $\alpha = x,y$. It is important to note that, as in the case of biquadratic symmetric exchange, these antisymmetric terms of the form $\sim [({\bf s}_i\times{\bf s}_j) \cdot {\bf \hat z}]^2$ can be deduced purely from symmetry arguments applied to a Hamiltonian with only spin degrees of freedom. \cite{walker}   In the case of CuFeO$_2$, ME coupling provides for one likely  microscopic mechanism.  The impact of $\mathcal{H}_{CP}$ on the evolution of spin states with increasing applied field is demonstrated below.  

\section{ Model Hamiltonian}

The full model Hamiltonian used here to describe the sequence ordered states of CuFeO$_2$ incorporates all of the effects described above.  Weak axial anisotropy is introduced in the form of the exchange (two-site) type $J_z({\bf r}_{ij})$ with possible physical origins from $Fe^{3+}$ defects \cite{whangbo06} or certain forms of antisymmetric superexchange exchange. \cite{shekhtman93} Anisotropy is also included in the biquadratic exchange term.  For simplicity, anisotropy of the single-ion type is omitted from the present analysis. With these considerations the Hamiltonian can be written as
\begin{eqnarray}
\mathcal{H} =  \mathcal{H}_{J} +  \mathcal{H}_{CP} +  \mathcal{H}_{G} + \mathcal{H}_{J_z} + \mathcal{H}_{G_z} - {\bf H} \cdot \sum_i {\bf s}({\bf r}_i)
\end{eqnarray}
where the first three terms are given above, bilinear exchange anisotropy is of the form 
$\mathcal{H}_{J_z} =  \sum_{ij} J_z({\bf r}_{ij})s^z_i s^z_j$ 
and biquadratic exchange anisotropy is given by 
$\mathcal{H}_{G_z} = -\sum_{ij}G_z({\bf r}_{ij}) [s^z_i  s^z_j]^2$.
The last term is the Zeeman coupling to a magnetic field ${\bf H}$ taken here to be applied along the hexagonal $c$ axis.  
In addition to the four quadratic Heisenberg exchange interactions  $J_1, J_2, J_3$, and $J'$, NN in-plane and inter-plane magnetoelectric coupling $C_1$ \cite{plumer91b} and $C'$, as well as NN in-plane and inter-plane biquadratic exchange couplings $G_1$ and $G'$ are included in the model calculations.  The impact of modifications to the in-plane exchange coupling due to the monoclinic lattice distortion at low temperature is accounted for in a model proposed in Ref. [\onlinecite{terada06}].  This effect was explored within the present model and was not found to substantially alter the qualitative results presented above.  

As a prelude to a discussion of the specific model Hamiltonian, and complimentary to the results of Mekata {\it et al.} \cite{mekata93}, the effect of interplane exchange coupling on stabilizing the P4 ground state phase is demonstrated here.  For this purpose, the Ising model is adequate.   The quadratic exchange contributions to the Hamiltonian are written as
$\mathcal{H}_J = \sum_{ij} J({\bf r}_{ij}){\bf s}_i\cdot {\bf s}_j$ 
where $J({\bf r}_{ij}) > 0$ for AF coupling and the sum is over sites within, as well as between, the $ABC$ stacked triangular layers which form the hexagonal crystal representation of the rhombohedral structure. Ground-state phases of Ising spins ${\bf s} || {\bf {\hat z}} || {\bf {\hat c}}$ with non-zero $J_1>0, J_2, J_3>0$ and $J'>0$ were determined analytically assuming the spin structures described by Mekata {\it et al.} \cite{mekata93} and verified numerically using the local-field method of Walker and Walstedt. \cite{walker2} Figure 1 demonstrates that long-period P4 and P8 spin configurations are stabilized by interplane exchange even in the absence of  $J_3$ but that third-neighbor in-plane coupling does serve to enhance the stability of these structures.   

\begin{figure}[h]
\includegraphics[width=0.4\textwidth]{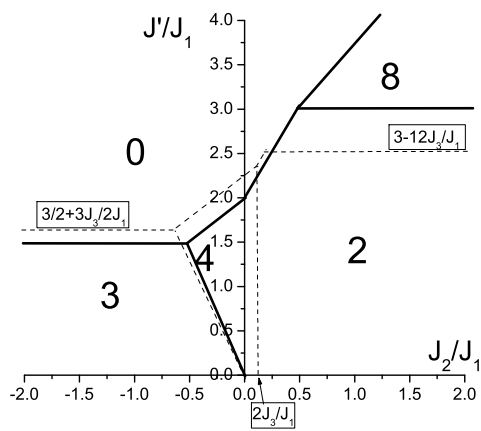}
\caption{\label{one} Ground state phases of the Ising model on a rhombohedral AF $J_1>0$ illustrating the effect of interplane coupling $J'>0$ on stabilizing longer-period structures described by Mekata {\it et al.}.\cite{mekata93}  Solid curves correspond to $J_3=0$ and broken curves $J_3 > 0$.}
\end{figure}

With only weak axial anisotropy, an applied magnetic field tends to destabilize a linear AF spin configuration with $s_i^z || {\bf \hat z}$.   An illustration of the effect of biquadratic symmetric exchange to enhance the stability of axially ordered states is shown in Fig. 2.  Using the model described above the evolution $(S^z)^2 = (1/N) \sum_i (S_i^z)^2$ (an average over all $N$ sites) in the P4 phase with increasing $H$ shows clearly this enhancement. 

\begin{figure}[ht]
\includegraphics[width=0.4\textwidth]{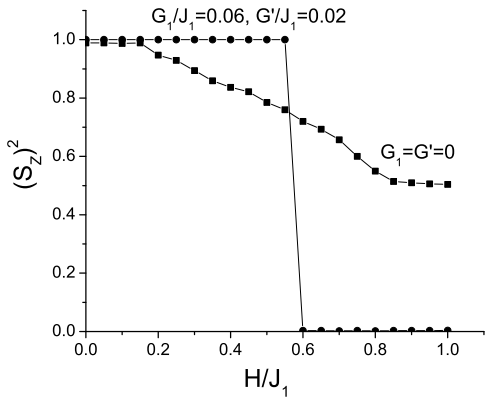}
\caption{\label{two} Effect of biquadratic exchange on the axial component of spins in the P4 state averaged over all sites with increasing magnetic field.  Model Hamiltonian (3) was used with 3\% anisotropy and $J_1=1, J_2=0.33, J_3=0.3, J'=0.4, C_1=C'=0$.}
\end{figure}

\section{Application to C\lowercase{u}F\lowercase{e}O$_2$}
 
The set of parameters required to reproduce qualitatively the sequence of field-induced spin configurations observed in CuFeO$_2$ is not unique. Some guidance can be found by estimation of exchange and anisotropy strengths based on the saturation and spin-flop fields, respectively, as done by Petrenko {\it et al.} \cite{petrenko05} Using their approach and accounting for the six NN in-plane interactions, leads to $J_1 \sim 3K$, with anisotropy being only a few percent of this value.    For simplicity, all of the anisotropy constants were set to be 3\% of their corresponding isotropic coupling strengths, i.e., $J_{kz}=0.03J_k ($k=1,2,3$), J'_z=0.03J', G_{1z}=0.03G_1, G'_z=0.03G'$.  All parameters are normalized to the NN in-plane exchange by setting  $J_1=1$. Other parameters of the model were chosen using guidance from previous model results and to yield a semi-quantitative representation of the unique sequence of field-induced phase transitions observed in CuFeO$_2$.
 
Ground state phases which minimize the full energy expression (3) were calculated numerically using the method of Walker and Walstedt \cite{walker2} with $ABC$ stacked triangular layers of dimensions $L \times L \times M$ , where $M=1$ corresponds to a unit cell containing all three layers. Periodic boundary conditions were imposed and spin configurations with in-plane periodicities ranging from L=1 - L=18 and up to M=4 unit cells along the $c$ axis were considered.  Absolute energy minima were found by comparing results among calculations using up to 50,000 different random initial spin configurations.  With this finite-size method, in-plane IC states characterized by $(q,q)$ modulations can be not be modeled directly.  However, such states can be deduced by demonstrating a slight reduction in energy as $L$ increases if $q \simeq n/L$ where $n$ is an integer.  For example, if the model energy is a minimum for an IC phase with $q \simeq 4.5$, the method will give a lower energy with $L=9$ ($n=2$) than either $L=4$ or $L=5$. 

Sets of parameter values which give the correct series of phases were determined partially through trial and error.  One such choice is $J_2=0.33, J_3=0.3, J'=0.4, C_1=0.3, C'=0.1, A_P=1, G_1=0.06, G'=0.02$. The ground state energies as a function of field corresponding to selected lower-energy periodicities $(L,M)$ are shown in Fig. 3 for this set of values.  The lowest energy states as $H$ increases follows the observed sequence (4,2)-(IC,2)-(5,2)-(3,1), with transitions at approximate values of $H_{c1} \simeq  0.75, H_{c2} \simeq 1.75$ and $H_{c3} \simeq 3.0$, following the notation of Ref.[\onlinecite{terada06}].  Saturation is found to occur at $H_{c5} \simeq 13.5$.  The phase labeled IC was deduced to be incommensurate since in this field regime L=4, 5, 9 and 14 states have nearly identical energy, with L=9 being the lowest.  This conclusion is consistent with the observed value of $q \simeq 2/9$ \cite{mitsuda00}.  Numerical accuracy of the method is limited at larger values of L. Note that the L=4, 9 and 14 phases have a simple periodicity of 2 along the c-axis ($q_z=1/2$, representing six triangular layers).  Additional Fourier components $q_z$ are also present in the [5,2] state. These characterizations are consistent with observed  neutron diffraction data. \cite{mitsuda00,petrenko00,nakajima07}  At higher field values, the [3,1] phase is found with $M=1$ (only three layers are required), followed by  saturated ferromagnetic $S_i^z || {\bf c}$ order.  Although substantial values of longer-range exchange interactions $J_2$ and $J_3$ are required to yield the correct sequence of phases, they are significantly reduced by inter-plane coupling from those assumed in Refs.[\onlinecite{mekata93,ajiro94,fukuda98}].  

\begin{figure}[ht]
 \includegraphics[width=0.5\textwidth]{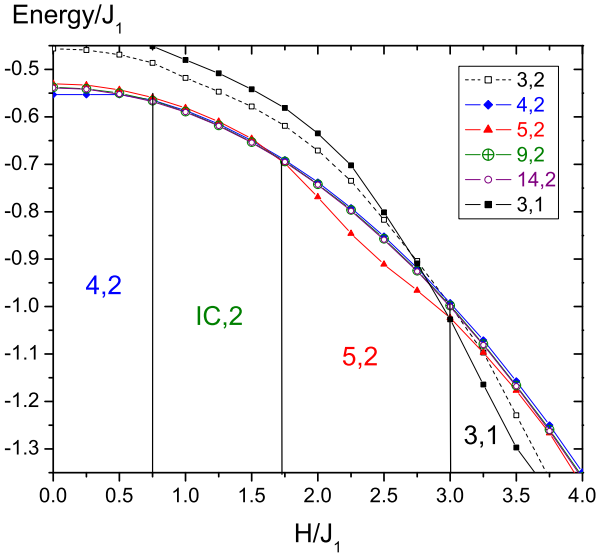}
    \caption{\label{three} Lower energy phases (L,M) as a function of magnetic field from the model Hamiltonian (3) using parameters $J_1=1, J_2=0.33, J_3=0.3, J'=0.4, C_1=0.3, C'=0.1, A_P=1.0, G_1=0.06, G'=0.02$ with 3\% axial anisotropy. }
 \end{figure}

Figure 4 shows the site averaged  $(S_z )^2$ with increasing $H$ associated with the phases which minimize the energy.  Spins in the (4,2) state are well aligned along the $c$ axis.  In the IC phase (represented by the (9,2) state) ${\bf S} \perp {\bf \hat c}$. In the (5,2) state, moments are aligned mostly along the $c$ axis but are slightly less oriented in (3,1) state until the field is close to $H=4$.   The spin reorientations at the boundaries (4,2)-IC and IC-(5,2) thus each represent spin-flop transitions.  All transitions are first order. In the IC phases, planar components $s_x \perp s_y$ have equal magnitude and the biquadratic antisymmetric exchange coupling stabilizes a helically polarized spin structure.  It is only in the IC state that an induced electric polarization ${\bf P} \perp {\bf \hat c}$ occurs. \cite{kimura06}

\begin{figure}[t]
 \includegraphics[width=0.5\textwidth]{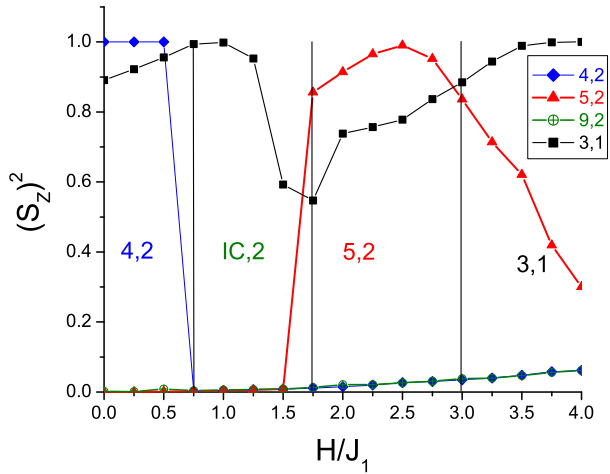}
    \caption{\label{four} Site averaged values of $(s_i^z)^2$ corresponding to the lowest energy states from Fig. 3.}
 \end{figure}

Figures 5-7 illustrate the spin structures in the triangular layers at representative field values for each of the ordered states. The high-field (3,1) phase ($\uparrow \uparrow \downarrow$) has a similar structure for each trilayer.  In the zero field (4,2) state ($\uparrow \uparrow \downarrow \downarrow$) spins on subsequent trilayers alternate in direction. In the  (5,2) phase, each of the six triangular layers in the magnetic unit cell are of the form ($\uparrow \uparrow \uparrow \downarrow \downarrow$) but the relation between spins on alternating trilayers $m=1$ and $m=2$ is more complicated, as found by Mekata {\it et al.}\cite{mekata93}  In the IC phase represented by the (9,2) structure of Fig. 7, spins lie in the triangular planes and rotate from site to site with a non-uniform angle.    
 
 \begin{figure}[t]
 \includegraphics[width=0.3\textwidth]{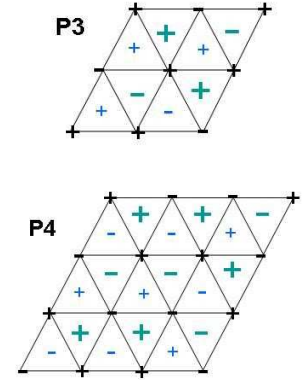}
    \caption{\label{four} Schematic representation of three triangular layers (distinguished by color and font size) of the (3,1) and (4,2) phases where $+$ and $-$ symbols denote spin vectors along the c-axis. In the (4,2) phase, all spin vectors are reversed in the subsequent trilayer ($m=2$).   }
 \end{figure}
 
 \begin{figure}[t]
 \includegraphics[width=0.35\textwidth]{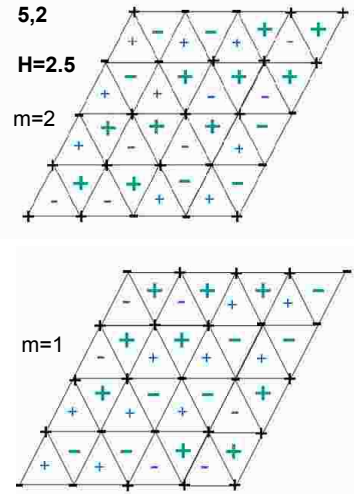}
    \caption{\label{four} Schematic representation (as in Fig. 5) of spins on the two sets of stacked trilayers characterizing the phase (5,2) at $H=2.5$.}
 \end{figure}
 
 \begin{figure}[t]
 \includegraphics[width=0.45\textwidth]{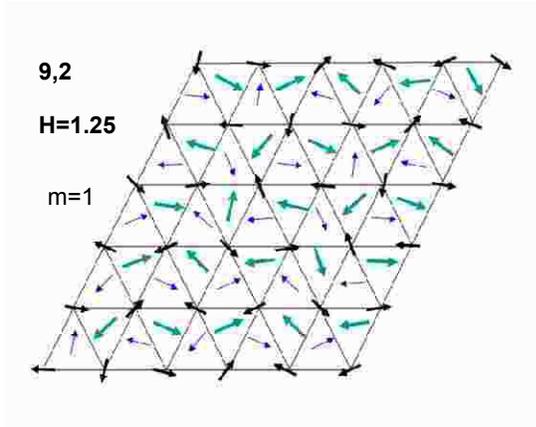}
    \caption{\label{four} An illustration of spin vectors (in the plane) on part of the lattice representing (9,2) magnetic structure at $H=1.25$. All spins are reversed in the subsequent tri-layer ($m=2$). }
 \end{figure}

Model results on the evolution of the magnetization $M_z=(1/N) \sum_i S_i^z$ with applied field is shown in Fig. 8 and may be compared with corresponding experimental data from Ref. [\onlinecite{terada06}].  Key features of the data are reproduced by the model such as $M_z=0$ in the (4,2) state, increasing magnetization with field strength in the IC phase, plateaus at 1/5 and 1/3 in the (5,2) and (3,1) states, respectively.  Flatter plateau regions, as seen experimentally, can result from the present model by increasing the strength of the anisotropy or biquadratic symmetric exchange (see Discussion below).   Note also that $M_z$ begins to deviate from the 1/3 plateau at fields above $H_{c4} \simeq 4.5$.  This represents the transition to the canted (3,1) state proposed in Ref.[\onlinecite{terada06}].  The five critical fields resulting from the present analysis may be compared with the experimental values  7 T, 13 T, 20 T, 34 T, and 70 T by multiplying the model field $H$ by a factor $(J_1 S/g \mu_B) \simeq 5.6 T$.  The resulting model critical-field values, 4.2 T, 9.8 T, 17 T, 28 T and 75 T, are in fair agreement with those deduced from the data, especially since no particular effort was made to adjust parameters for this purpose.  

\begin{figure}[t]
 \includegraphics[width=0.5\textwidth]{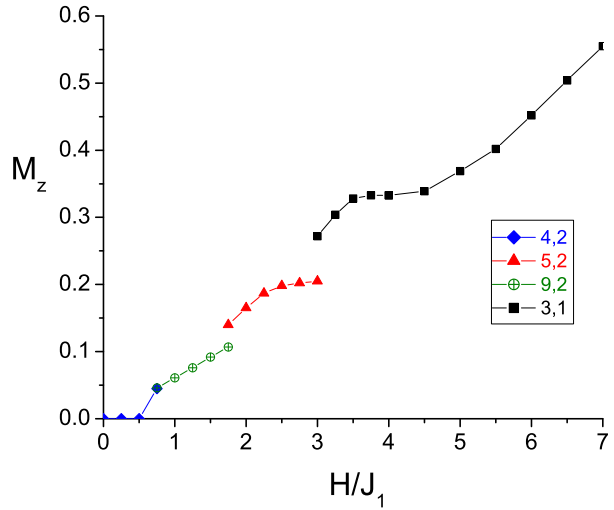}
    \caption{\label{four} Site averaged values of $(s_i^z)^2$ corresponding to the lowest energy states from Fig. 3.}
 \end{figure}

\section{Discussion and Conclusions}

A crude estimate of the contribution to biquadratic antisymmetric exchange interaction due to magnetoelectric coupling can be found using data on the dielectric constant $\epsilon \simeq 20$ and polarization $P \simeq 400 \mu C/m^2$ from Ref. [\onlinecite{kimura06}]. With three $Fe^{3+}$ ions per unit cell \cite{whangbo06}, the corresponding energy per magnetic ion is a very small $E_P/N \sim  P^2/(\epsilon_0 \epsilon) \simeq 0.003K$.  Equating this to the antisymmetric exchange energy $E_{CP} \sim C_1^2/(8A_P)$, yields the crude estimate $C_1  \sim 0.1$, somewhat smaller than the value used in the present analysis $C_1  = 0.3$.  Recall, however, that the contribution $\mathcal{H}_{CP}$ to the spin Hamiltonian can be deduced from general symmetry considerations and can in principle have a variety of microscopic origins, possibly stronger than from the magnetoelectric effect, such as higher order hopping analogous to symmetric biquadratic exchange.\cite{takahashi77,shekhtman93} (Note that an analysis of the present model with both stronger antisymmetric exchange and stronger anisotropy leads to an increase in all the transition fields as well flatter magnetization plateaus). Spin-flop transitions occur when the Zeeman energy $E_Z=-{\bf S} \cdot {\bf H}$ equals the anisotropy energy $E_A$ of a magnetic system, $\Delta E = E_z - E_A = 0$.  Even very small perturbations can affect the system in the transition region, particularly for highly frustrated systems. In the present case, the small antisymmetric biquadratic exchange interaction stabilizes the IC phase. This point is also demonstrated by considering an analysis of the model Hamiltonian in the absence of the term $\mathcal{H}_{CP}$.  The evolution of lowest energy states with $C_1 = C' = 0$ is shown in Fig. 9, with all other parameters set as above.  The IC phase as characterized above is never stabilized, even when considering a wider range of values for the other model parameters.   

\begin{figure}[ht]
 \includegraphics[width=0.5\textwidth]{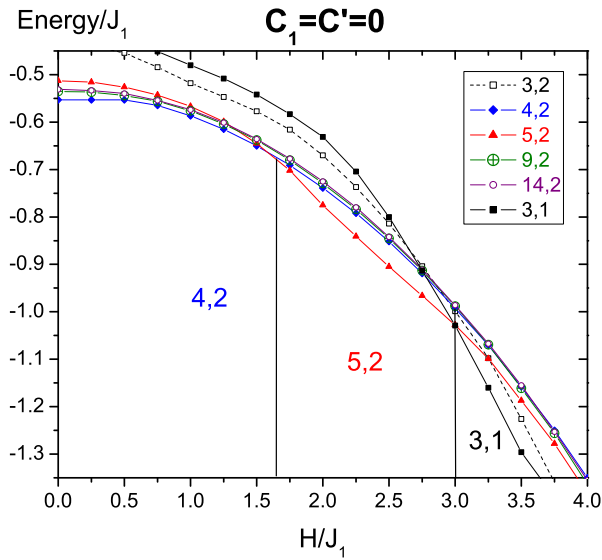}
    \caption{\label{three} Lower energy phases as in Fig. 3 calculated with no biquadratic antisymmetric exchange.}
 \end{figure}
 
Preliminary analysis of a Landau-type free energy \cite{plumer88} based on a molecular-field treatment of the  Hamiltonian    (3) indicates that the proposed model captures essential features of the magnetic-field -- temperature phase diagram. In this   formalism, the free energy is written as a functional of the spin density ${\bf s}( {\bf r}) = {\bf m} + {\bf S} e^{i{\bf Q \cdot r}} + {\bf S^*} e^{-i{\bf Q\cdot r}}$ where ${\bf m}$ is the uniform magnetization and ${\bf S}$ and ${\bf Q}$ represent polarization and wave vectors, respectively, of the modulated spin structure.  In zero applied field, the observed sequence of transitions from paramagnetic to linearly polarized IC to P4 states is reproduced.  The stability of the linearly polarized P4 phase is associated with fourth-order zero-field Umklapp terms of the form $[{\bf S} \cdot {\bf S}]^2 \Delta_{4{\bf Q},{\bf G}}$, where ${\bf G}$ is a reciprocal lattice vector.  Similarly, sixth-order terms $[{\bf m} \cdot {\bf S}][{\bf S} \cdot {\bf S}]^2 \Delta_{5{\bf Q},{\bf G}}$ enhance the stability of the field-induced linearly polarized P5 state. The higher field period-3 phase occurs due to terms of the form $[{\bf m} \cdot {\bf S}][{\bf S} \cdot {\bf S}] \Delta_{3{\bf Q},{\bf G}}$. The helically polarized IC phase is characterized by ${\bf S} \perp {\bf S^*}$ \cite{plumer91b} is a result of contributions arising from $\mathcal{H}_{CP}$. Detailed results will be reported elsewhere. 

The present work demonstrates that the complex sequence of ordered phases observed in CuFeO$_2$ stabilized by increasing magnetic field strength is a consequence of a high degree of frustration due to a multitude of competing interactions.  The usual basal-plane period-3 spin structures associated with the STAF are destabilized by longer-ranged intra-plane as well as inter-plane quadratic exchange interactions.  Weak axial anisotropy is found to be enhanced by biquadratic symmetric exchange, likely due to magnetoelastic coupling.  Magnetoelectric interactions through spin-orbit coupling provides for one mechanism that gives rise to a new type of biquadratic {\it antisymmetric} exchange term which serves to stabilize the helically polarized spin-flop phase.  The model Hamiltonian proposed here will serve as the foundation for future work focused on magnetoelastic effects, spin excitations \cite{terada07} and finite-temperature effects.  Many of the features of the present model are likely relevant to other recently examined magnetoelectric STAF's. \cite{lawes} 

%




I thank O. Petrenko, Y. Ren, G. Quirion and S. Nagler for enlightening discussions. This work was supported by the Natural
Science and Engineering Research Council of Canada (NSERC) and the Atlantic Computational Excellence Network (ACEnet).

\end{document}